\documentstyle[epsf,fleqn,aps]{revtex}

\def\Section#1{}

\newcommand{\be}{\begin{eqnarray}}
\newcommand{\ee}{\end{eqnarray}}
\newcommand{\beq}{\begin{equation}}
\newcommand{\eeq}{\end{equation}}
\newcommand{\xx}{\begin{eqnarray*}}
\newcommand{\yy}{\end{eqnarray*}}

\def\beq{\begin{equation}}

\def\eeq{\end{equation}}

\def\bea{\begin{eqnarray}}

\def\eea{\end{eqnarray}}

\def\age{\,\raise2pt\hbox{$\mathop{>}\limits_{\raise 2pt

\hbox{$\sim$}}$}\,}

\def\ale{\,\raise2pt\hbox{$\mathop{<}\limits_{\raise 2pt

\hbox{$\sim$}}$}\,}

\setbox122 = \hbox{1}

\def\id{\rlap{1}\rlap{\kern 1pt \vbox{\hrule width 4pt depth 0 pt}}

        \rlap{\kern 4 pt \hbox{\vrule height \ht122 depth 0 pt}}

           \hskip\wd122}

\begin{document}
\tolerance 50000
\twocolumn[\hsize\textwidth\columnwidth\hsize\csname
@twocolumnfalse\endcsname

\title{From classical to quantum  Kagom\'e antiferromagnet in a magnetic field}
\author{D.C.\ Cabra$^{1,2}$, M.D.\ Grynberg$^{1}$,
        P.C.W.\ Holdsworth$^{3}$ and P.\ Pujol$^{3}$}

\address{$^{1}$Departamento de F\'{\i}sica, Universidad Nacional de la Plata,
               C.C.\ 67, (1900) La Plata, Argentina.\\
$^{2}$Facultad de Ingenier\'\i a, Universidad Nacional de Lomas de
      Zamora, Cno. de Cintura y Juan XXIII,\\
      (1832) Lomas de Zamora, Argentina.\\
$^{3}$Laboratoire de Physique,
      ENS Lyon, 46 All\'ee d'Italie, 69364 Lyon C\'edex 07, France.\\}

\maketitle

\vspace{.5cm}

\begin{abstract}
We study the magnetic properties of the Kagom\'e antiferromagnet
going from the classical limit to the deep quantum regime of spin
$1/2$ systems. In all the cases there are special values for the
magnetization, $1/3$ in particular, in which a singular behavior
is observed to occur in both the classical and quantum cases. We
show clear evidence for a magnetization plateau for all $S$, in a
wide range of $XXZ$ anisotropies and for the occurrence of {\it
quantum} order by disorder effects.

\vskip 0.5cm

PACS numbers: 75.10.Jm, \,75.10.Nr,\, 75.60.Ej.

\end{abstract}


\vskip -0.2cm \vskip2pc]


Spin systems in low dimensions have become an intense area of
research in the last few years due to their possible relevance to
high-$T_c$ cuprates but also in their own right due to their very
special properties \cite{review}. One of the issues that attracted
 attention,  both from experimentalists and theorists, is the
appearance of plateaux in the magnetization curves in both 1D
\cite{OYA,nos,Totsuka} and 2D \cite{Hone,Oni,MT,Muller,MJG}
quantum spin systems. Another has been the study of strongly
frustrated systems with highly degenerate classical ground state
structure \cite{Moessner} giving disordered, spin liquid behavior
at low temperature. The spin liquid behaviour can be removed, or
partial removed by degeneracy lifting, through thermal, or quantum
fluctuations; the phenomenon known as order by disorder
(OBD)~\cite{Villain,Shender}. Examples are the so-called $J_1-J_2$
model on a square lattice~\cite{SZ} and the Kagom\'e Heisenberg
antiferromagnet (AF), where thermal OBD leads to a nematic spin
liquid state~\cite{CHS,SH,RB}. The situation is more complicated
when one considers quantum spins defined on such lattices, and the
role of quantum fluctuation in the selection of some particular
ground states (GS's). In the particular case of the Kagom\'e
lattice \cite{AS}, state of the art numerical computations
\cite{Phillipe} seem to indicate that no nematic order is observed
and that the GS is a rather complicated spin liquid like state,
with a spin gap but with many low lying spinless states.

Recent studies of the Kagom\'e AF  in the presence of a magnetic
field have shown that the value of $1/3$ for the normalized
magnetization is a special point bringing into play both OBD
effects and the presence of a magnetic plateau, an effect also
present for the triangular lattice \cite{Hone,CG}. In the
classical case~\cite{ZHP,Z}, spin wave analysis and Monte Carlo
simulation show the order by disorder selection of a collinear
spin-liquid state together with a magnetic anomaly and a pseudo
magnetic plateau. In the quantum case, numerical evidence from the
exact diagonalization of small $S = 1/2$ clusters show the
existence of a magnetic plateau~\cite{Hida}. Such behavior has
already been observed in the $J_1-J_2$ square lattice, where
similar phenomena, based on a collinear order of the spins for
particular values of the magnetization, occurs \cite{ZHP}.

A natural question one can pose is therefore: is there a
connection between the existence of soft modes in the classical
case and the appearance of magnetization plateaux in the quantum
system ? Further; is the magnetic plateau accompanied  by quantum
order by disorder selection of a reduced manifold of states ? In
this paper we have addressed these questions
 using a standard spin wave analysis,
Lanczos diagonalization of finite clusters and a 2D Jordan-Wigner
approach. We study the $XXZ$ AF on the Kagom\'e lattice, where in
general no soft modes are present. Our results show the
persistence of the plateau at $1/3$ beyond the symmetric XXX
point, and even the existence of a plateau for classical spins, in
the easy axis case. This suggests that the existence of the soft
modes is not a necessary condition for the existence of a plateau
in the quantum case, although it does seem to extend the plateau
region away from the Ising limit of $XXZ$ AF. The quantum ground
state does indeed have a large projection onto a collinear state
indicating a quantum order by disorder effect.

We define the $XXZ$ AF Hamiltonian \beq \!\!\!\! H = J
\sum_{<i,j>} \left(S_i^x S_j^x + S_i^y S_j^y + \Delta S_i^z
S_j^z\right) - \vec h \cdot  \sum_i \vec S_i, \label{1} \eeq where
$\vec S_i=(S_i^x,S_i^y,S_i^z)$ is a vector of unit length, $J
>0$ is the exchange constant and $0 < \Delta < \infty$ the
``$XXZ$" anisotropy parameter. Specific values, $\Delta = 1$ and
$\Delta = 0$ define the $XXX$ and $XX$ models respectively.

Considering first the $XXX$ case and defining $\vec S_p =
\sum_{i=1,2,3} \vec S_i$ as the total moment per triangular unit,
the Hamiltonian for $N$ spins can be written as
\beq H = -NJS^2 + \sum_{p=1}^{2N/3}  \left( {J\over{2}}
 \vec S_p^2 - {1\over{2}}\vec h \cdot \vec S_p\right) \ .
\eeq
Minimizing with respect to $\vec S_p$ gives $\vec S_p = \vec
h/(2J)$ and a net GS magnetization $\vec M = \vec h/(6JS)$, for $
|\vec h| < h_c = 6JS$. Above this field the spins are
ferromagnetically aligned. In zero field this condition leads to a
rigid coplanar spin configuration for each triangular plaquette,
but the open nature of the Kagom\'e lattice allows freedom in the
orientation of the triangular spin planes and a highly degenerate
continuous GS manifold. Spin wave analysis shows that a branch of
modes for fluctuations out of a discrete, but macroscopically
large subset of collectively coplanar ground states become soft,
leading to their thermal selection~\cite{CHS,SH,RB,RCC}.

For non-zero magnetic field, a generic classical ground state for
a single plaquette is not necessarily coplanar, in contrast to the
zero field case~\cite{MC}. Selection of a collective coplanar
state therefore requires selection at the level of each triangle,
as well as between triangles.

This fact weakens, but does not destroy the coplanar
selection~\cite{SH}. For small field Zhitomirsky has proposed that
a discrete set of coplanar states are selected with, on each
triangle one spin, $A$, antiparallel to the field and spins $B$
and $C$ straddling the field direction in a ``$Y$'' configuration.
As in zero field, these states have soft modes associated with out
of plane spin fluctuations and corresponding line defects in real
space, allowing rotation of spins $B$ and $C$ about $A$.

As $ h \rightarrow h_c/3$ the two canted spins close up onto the
field direction, forming collinear ``up-up-down'' (UUD) spin
configurations on each triangle, with magnetization $M = 1/3$.
This special point has zero modes for in plane, as well as out of
plane spin  fluctuations~\cite{ZHP,Z}. The large number of soft
modes leads to strong thermal selection of the UUD spin
liquid~\cite{ZHP,Z}.

To describe the fluctuations over the coplanar states near
$h_c/3$, we choose a coordinate system oriented with respect to
local spin directions \cite{CHS}. At each site we choose right
handed axes in spin space with the ``$z$'' axis parallel to the
spin and all ``$x$" axes mutually parallel, that is, perpendicular
to the spin plane. The spin orientations are therefore
parameterized by $\vec{S}_i = (\epsilon_i^x,\epsilon_i^y, 1 -
\alpha_i)$, with $\alpha_i$ determined from $|\vec{S}_i| =1$ and
the quadratic part of the Hamiltonian reads
\beq \label{H2} 1/2~ \sum_{i,j}~\left[ M^x_{i,j} \epsilon_i^x
\epsilon_j^x + M^y_{i,j}\epsilon_i^y \epsilon_j^y \right]. \eeq
The only non-zero elements of the matrices $M^x$ and $M^y$ are the
diagonal and nearest-neighbors terms. In particular, $M^x$ entries
are $M^x_{i,i} = 1$ and $M^x_{i,j} = 1/2$ if $i$ and $j$ are
nearest-neighbors. While $M^x$ turns out to be independent of $h$
and the particular GS we are considering, this is not the case for
$M^y$.

The analysis of the ``$x$" sector is given in reference
\cite{CHS}. Even for disordered states the fundamental excitations
are propagating modes, rather than the localized line defects that
provide an alternative description of the soft
fluctuations~\cite{RCC}. One can diagonalize (\ref{H2}) in Fourier
space finding hidden branches of magnon-like excitations, as far
as magnetic scattering techniques are concerned: an optical
branch, an acoustic branch and a soft branch extending over the
entire Brillouin zone. This branch is responsible for the
selection of coplanar states in zero field and is present for all
field values \cite{SH,Z}. At the point $h/h_c=1/3$ and for
fluctuations around an UUD state the ``$y$'' sector also acquires
a soft branch, independently of the GS.
This fact plays in favor of a selection of the UUD states and
leads to the UUD ``spin-liquid'' predicted and observed in
reference \cite{ZHP}.


\hbox{%
\vspace {1.2cm}}
\begin{figure}
\hbox{%
\epsfxsize=3.2in \hspace{0.0cm} \epsffile{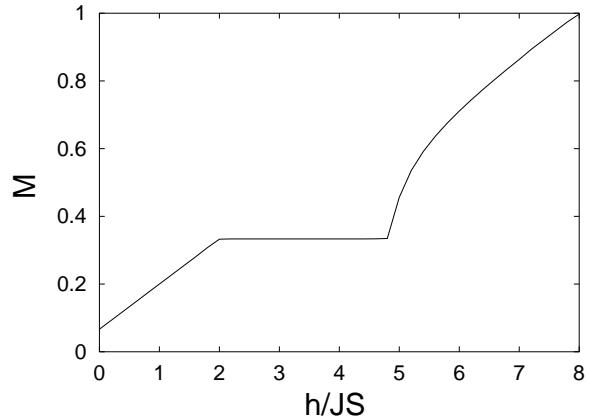}} \vskip 0.5cm

\caption{Zero temperature magnetization curve, for classical
spins. $\Delta = 1.5$. } \label{fig0}
\end{figure}


\hbox{%
\vspace {0.0cm}}
\begin{figure}
\hbox{%
\epsfxsize=2.2in \hspace{1.0cm} \epsffile{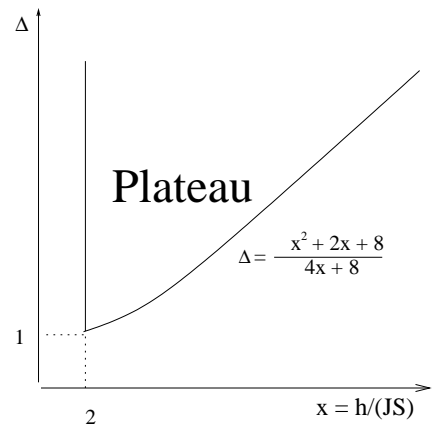}} \vskip 0.5cm

\caption{Schematic classical phase diagram in the $h$, $\Delta$
plane.} \label{fig0-bis}
\end{figure}

The singular entropy of fluctuations driving the UUD spin liquid
also leads to a magnetic anomaly and the beginnings of a magnetic
plateau. Close to $h = h_c/3$ the growing entropic contribution to
the free energy pulls the minimum away from the ground state
energy and towards the UUD point. This pulls the magnetization
above the linear curve $M = h/h_c$ as one approaches $h_c/3$ from
below and pulls it below as one approaches from above. However the
spin wave analysis  gives no plateau, rather it gives a continuous
relaxation away from linear behaviour and a point of inflection at
$h=h_c/3$. A remaining open question is whether nonlinear
corrections can give a discontinuity and a finite plateau, as
proposed in~\cite{Z}.

For $\Delta >1$ and in zero field the ground states remain highly
degenerate and there are soft modes for out of plane fluctuations
about collectively coplanar states with moment $M = (\Delta
-1)/3(\Delta +1)$~\cite{BGR}. Adding a small field along the $z$
axis does not lift the degeneracy or remove the soft modes. The
magnetization increases linearly until $M =1/3$ at $h=2JS$,
independently of $\Delta$. For higher fields,  there is a plateau
in the magnetization, at $ M= 1/3$, continuing out to $h = (2
\Delta - 1 + \sqrt{4\Delta^2+4\Delta-7}) JS$ where it starts to
grow again, until reaching saturation at $h = (2 + 4 \Delta) JS $.
The plateau occurs because, in this field region the field and
exchange contributions to the Hamiltonian both vary quadratically
with deformation away from an UUD configuration, leading to a
linear instability. As a result the macroscopic, but discrete set
of UUD states is selected energetically rather than entropically
and the field must exceed $h=2JS$ by a finite amount, before the
ground state manifold leaves these configurations. The in-plane
soft modes at $M = 1/3$ are however a property of the $XXX$ point
and are removed for $\Delta
> 1$. The lower boundary of the plateau tends to $M=0$
as one approaches the Ising limit, $\Delta
\rightarrow \infty$, and there is a zero temperature plateau
extending over the whole of the range $h < h_c$. The zero
temperature magnetization for the classical case, is shown in Fig.
\ref{fig0}, for $\Delta = 1.5$. The data were obtained from zero
temperature Monte Carlo, decreasing the field the starting value
$h=h_c$. The plateau region is clearly visible. For $\Delta <1$
and zero field, the out of plane fluctuations acquire a finite
stiffness~\cite{RCC}. In finite field the ground states are formed
by canting the spins out of the plane perpendicular to the field
direction. There are in general no soft modes and $M=1/3$ no
longer has any special significance.

Classically, there is therefore an extended region  in the
$\Delta, h$ phase diagram at low temperature where plateau
phenomena can be observed, extending from the point $\Delta=1$,
$h=2JS$ in a wedge, out to the line $\Delta=\infty, h < h_c$, as
is shown schematically in Fig. \ref{fig0-bis}. A natural question
to ask is therefore, do quantum fluctuations extend this region,
leading to a plateau over a larger region of parameter space?

In trying to understand this problem, we present here a brief
description of quantum fluctuations over coplanar states for
arbitrary values of $M$. Within a Holstein-Primakoff description
for the spin variables, where the quantization axis is chosen
according to the classical orientation of the spin described
above, the $XXX$ Hamiltonian can be written:
\beq H = H_0 + {S\over 2} ( H_2 + O( 1/\sqrt{S})) \eeq
where $H_2$ is a quadratic Hamiltonian of magnon creation and
annihilation operators. The $O(1/\sqrt{S})$ part contains higher
order terms in these operators. By expressing the creation and
annihilation operators in terms of the corresponding canonical
variable $Q$ and $P$, $H_2$ reads~\cite{HKB}:
\beq H_2 = 1/2 \sum_{i,j}~\left[ M^p_{i,j} P_i P_j + M^q_{i,j} Q_i
Q_j \right] \eeq
where $M^p$ and $M^q$ are the matrices for classical fluctuations
in the $y$ and $x$ sectors discussed above. There are therefore
$N/3$ $Q$ coordinates absent in $H_2$ with the corresponding
canonical variables $P$ commuting with $H_2$. This is the quantum
analog of the flat branch. It can be explicitly shown to exist for
the ordered coplanar GS with $q=0$ ordering vector, using a three
color boson picture in momentum space. In this case, by an
appropriate canonical transformation, $H_2$ can be written as:
\beq \!\!\!\!\!\!\!\!\!\!\! H_2 = \frac{1}{2}
\sum_{\vec{k},\mu}\left[ P_\mu (\vec{k}) P_\mu (\vec{-k}) +
\omega_\mu (\vec{k}) X_\mu (\vec{k}) X_\mu (\vec{-k}) \right]\,,
\eeq where $\mu$ is the color index and $\omega_3(\vec{k}) = 0
~\forall~ \vec{k}$. We see then the presence of the flat mode in
the quantum case, at this order in $1/ \sqrt{S}$, as a free
particle like spectrum.

Of particular interest are the UUD states, where $N/3$ $P$
coordinates are also absent from the Hamiltonian at this order. It
is obvious that higher order terms of the Hamiltonian are crucial
for understanding the spectrum of these excitations. In \cite{DS}
it was shown how the zero-point fluctuations of the fast variables
(here $i=1,2$) introduce (via the higher order terms) an effective
potential for the soft modes, apart from lifting the classical GS
degeneracy.
Following \cite{DS}, processes like the tunneling effect between
the, now discrete set of local minima can give rise to a finite,
and even gaped stiffness of the soft modes. What we should point
out in this example, is that the case $|M| = 1/3$ corresponds to a
different situation, where the effective dynamics of the soft
modes introduced by the interactions with the normal oscillators
will be radically different. We argue that, because of the absence
of this huge number of canonical variables in $H_2$ for the UUD
states, quantum fluctuations will indeed  play in favor of a
selection of these states, and to their stability as the magnetic
field is changed away from $h_c/3$. Such a selection, for a
magnetic UUD state, is also present for the triangular lattice,
due to zero-point fluctuations over an ordered classical ground
state~\cite{CG,KM}. Following this idea one could follow a
standard semiclassical expansion, and argue that if quantum
fluctuations select a "Y" coplanar configuration for $M < 1/3$ and
a quasi-collinear configuration for $M > 1/3$ as do thermal
fluctuations in the classical case \cite{Z}, the value of the
field at which $ M = 1/3$ is reached from above and below will be
renormalized at the quantum level. This scenario gives rise to a
plateau at $M=1/3$ even for $\Delta =1$. Indeed, taking for
example a $q=0$ structure, a straightforward numerical integration
of the free bosonic branches involved  either in the ``$Y$'' or
quasi-collinear Kagom\'e vacua, yields a finite plateau width.
While this estimation is based on a particular
ground-state ($q=0$ in this case), it is reasonable to assume a
similar behavior for any coplanar ground-state. A more subtle point concerns
the fact that this approach has to be taken with some care for the Kagom\'e
model, because the oscillator frequencies have a non-analytic behavior at
$h=2JS$. This can give rise to divergencies in the coefficients of the formal
power series and produce a non-analytic dependence on $S$ of the
width of the plateau.

Motivated by these semiclassical arguments in favor of a plateau
at $1/3$, we further analyze this issue using Lanczos
diagonalization and a Jordan-Wigner approach, which are more
appropriate in the deep quantum case \cite{Leche}. Let us first
focus attention on the exact GS's obtained from the
diagonalization of the clusters with periodic boundaries
schematized in Fig.\ 3. Using a Lanczos algorithm \cite{Lanczos}
applied through all $S^z$ subspaces, we observe a plateau at
$M=1/3$  as a function of magnetic field. Results for $N=18$ and
all 24-spin clusters are shown  in Fig.\ 4 for both $XXX$ and
$XXZ$ with $\Delta >1$.

\hbox{%
\vspace {-2.5cm}}
\begin{figure}
\hbox{%
\epsfxsize=3.4in \hspace{0.2cm} \epsffile{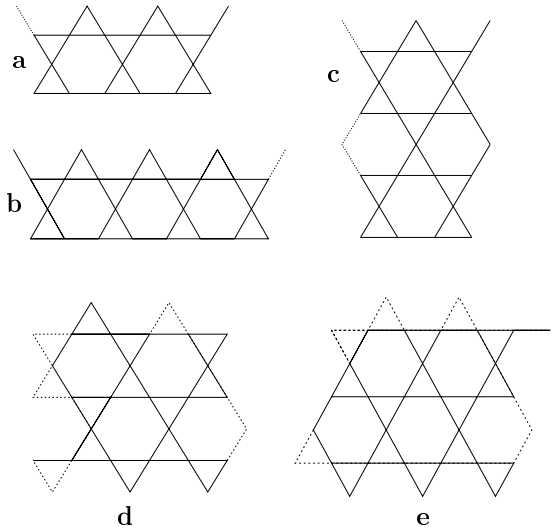}}
\vspace{-2.5cm} \caption{Schematic view of clusters considered in
exact diagonalizations. (a) 18 spins; (b), (c) and (d) 24 spins;
(e) 27 spins. Dotted lines indicate PBC, holding throughout all
boundaries. \label{fig3} }
\end{figure}

\begin{figure}
\hbox{%
\epsfxsize=4in
 \hspace{-1cm} \epsffile{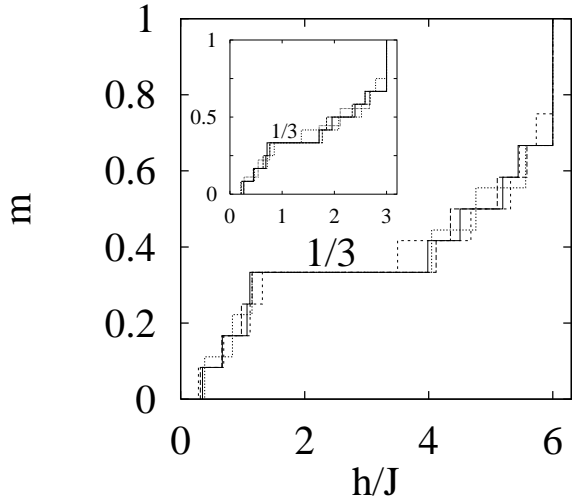}}
\vskip -1.5cm \caption{Magnetization curves for $XXZ$ clusters
$(\Delta = 2.5)\,$. Dashed, solid, short dashed, and dotted lines
denote respectively the results obtained for clusters (a), (b),
(c) and (d). The inset displays the corresponding results for
isotropic exchanges $(\Delta = 1)\,$. \label{fig4} }
\end{figure}

To check whether an UUD order actually occurs, we measured
three-point coplanar correlation functions of the form
$C_{\theta}(i,j,k) = \sum_{< i,j,k>} \langle \,\hat n^{\theta}_i\,
\hat n^{-\theta}_j\, (1-\hat n^z_k)\,\rangle\,$ on the $M = 1/3\,$
GS. Here, $<i,j,k>$ denotes a cyclic permutation around an
elementary triangle whereas, $\hat n^z = \sigma^+ \sigma^-$ is the
usual spin occupation number and $\hat n^{\theta}$ stands for the
local density fields along the $\theta$-axes forming the coplanar
``$Y$'' configuration. Specifically, after some elementary steps
it can be readily shown that
\begin{eqnarray}
\nonumber \hskip -0.7cm C_{\theta}(i,j,k) &=&
\frac{1}{4}\!\!\sum_{< i,j,k>}\!\! \left\langle\, \left[\;
-\sin^2\theta\; \sigma^x_i\,\sigma^x_j + (1+ \cos \theta
\,\sigma^z_i) \right.\right. \\ &\times& \left. \left.\, (1+ \cos
\theta \,\sigma^z_j)\, \right]\, (1-\hat n^z_k)\,\right\rangle\,.
\end{eqnarray}
It turns out that $C_{\theta}$ increases monotonically with
$\Delta$ and near the maximum depends very little
either on the cluster size, its particular shape,  or the
plaquette location. A large but unsaturated value of this
parameter at $\theta = 0\,$, as is shown in Fig. 5, is consistent
with the idea of a {\it collinear} UUD spin liquid allowing a
finite probability for configurations with defects.
\hbox{%
\vspace {-4.7cm}}
\begin{figure}
\hbox{%
\epsfxsize=4in \hspace{-0.5cm} \epsffile{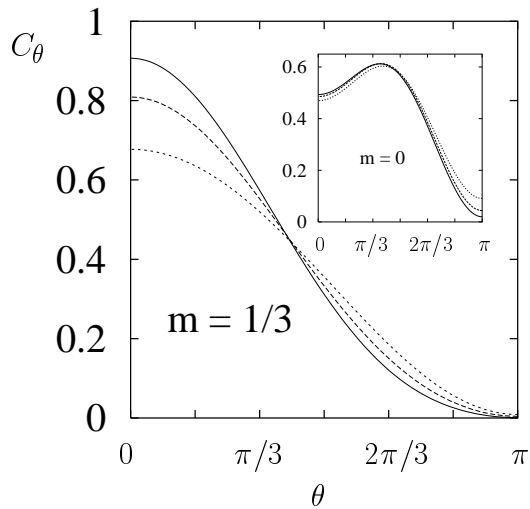}} \vskip -1.5cm
\caption{Triangular correlation functions measuring variations
around the coplanar "Y" configurations at $M = 1/3$ in cluster
(d). Solid, dashed and dotted lines refers to $\Delta = 2.5, 1,$
and $0$ respectively. For comparison, the inset shows the
corresponding situation a $M=0$, favoring a classical behavior.
\label{fig5} }
\end{figure}

One can follow this idea further by testing if the system selects
a single state or microscopic number of states making up the UUD
spin liquid. By sorting the probabilities of the ground state
components in real space, we find that the latter exhibit a
sharply peaked distribution, at the top of which there are a
number of UUD states. This is shown schematically in Fig. 6 for
cluster (e). The weight distribution becomes even sharper for
$\Delta > 1\,$ and broadens steadily upon decreasing the
anisotropy. Similar results were observed in the remaining
clusters displayed in Fig.\,3\, \cite{boundaries}. This ensemble
of results therefore gives a strong indication that there is
quantum order by disorder selection of the UUD spin liquid over
the plateau region for $\Delta = 1$. The narrowing for $\Delta >
1$ is evidence for the removal of non-collinear fluctuations  away
from the symmetric $XXX$ point, as one might expect from the
energetic arguments presented earlier.

\hbox{%
\vspace {-4.2cm}}
\begin{figure}
\hbox{%
\epsfxsize=4.2in \hspace{-1.4cm} \epsffile{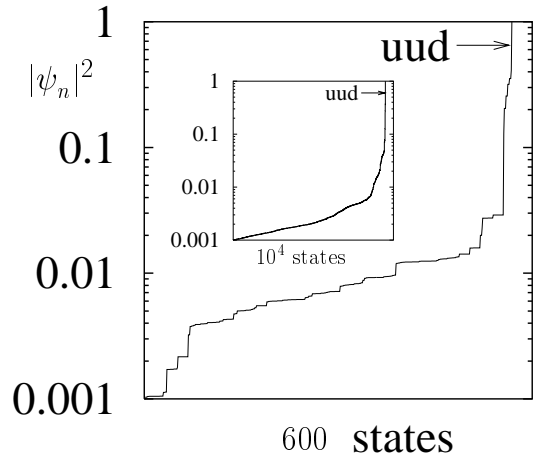}} \vskip -3cm
\caption{Ground state of cluster (e) for $M = 1/3\,$ and $\Delta =
2.5\,$. Real space states are sorted by increasing weights and
normalized to their maximum value (up-up-down configurations). The
inset exhibits the ground state components for $\Delta = 1\,$.
Only weights above $10^{-3}\,$ are shown. \label{fig6}}
\end{figure}

As a final step, we now study our system by means of a
generalization of the well known Jordan-Wigner transformation in
1D, which was originally proposed in \cite{Fradkin} and further
developed in \cite{2DJW,YWG,LRF}. It consists of a mapping of the
original spin variables into spinless fermion variables $c_i$ via
$S^-_{i}=c_i \exp i \sum_{j\neq i} c^\dagger_j c_j \arg
(\tau_j-\tau_i)$, where $\tau_i=x_i+iy_i$ is the complex
coordinate and $S^z_i=c^\dagger_i c_i -1/2$. The advantage of this
transformation is that it imposes exactly the single occupancy
constraint, but at the price of introducing a minimal coupling
with Chern-Simons gauge fields. One usually proceeds by a mean
field decoupling to end up with a problem of fermions on the given
lattice in a properly chosen magnetic background, which can be
solved using standard techniques. In this language, an external
magnetic field applied to the spin system enters as a chemical
potential for the spinless fermions. In the present case and
mainly motivated by the tendency of the AF Kagom\'e lattice to
disorder , we have chosen a homogeneous mean field, that is a
uniform fermion density $c^\dagger_j c_j \equiv \frac{1}{2} ( 1\,+
M  \,),\:\, \forall j\,$. We restricted the analysis to the case
$\Delta = 0\,$ where this calculation scheme becomes most
reliable. The resulting magnetization curve is shown in Fig.\ 7. A
clear plateau again shows up at $M=1/3$. Two other plateaux also
appeared at $M=0$ and $M=2/3$ which could be an artifact of the
mean field approximation and more work is needed to be conclusive
at this point.


\hbox{%
\vspace {-4.5cm}}
\begin{figure}
\hbox{%
\epsfxsize=4in \hspace{-1cm} \epsffile{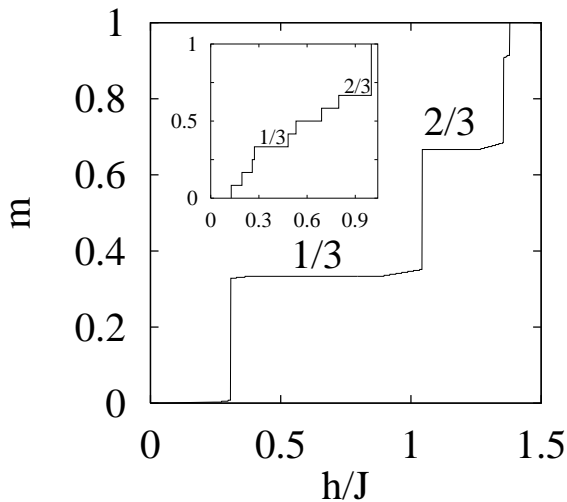}} \vskip -1.5cm
\caption{Mean field magnetization curves for $XX$ lattices, using
homogeneous fermion densities. For comparison, the inset shows the
exact magnetization curve of cluster (c) using $\Delta = 0\,$.
\label{fig7}}
\end{figure}

It is interesting to recall that the $1/3$  plateau is expected to
disappear for low enough $XXZ$ anisotropy in the triangular
lattice \cite{Hone}, while our mean-field results suggest that it
persists even in the $XX$ limit of the Kagom\'e lattice. This
result is again consistent with the idea that quantum fluctuations
stabilize a plateau over an extensive range of parameter space.
The Lanczos magnetization curve at $\Delta = 0\,$ yields no
conclusive evidence on this issue, though there is evidence that
the sharp weight structure of the GS components disappears.

To summarize, we have studied the magnetization behavior of both
classical and quantum  Kagom\'e AF using a variety of methods. In
all cases our results suggest that at least for $\Delta \ge 1\,$,
there is selection of an UUD state which is ultimately responsible
for the emergence of a magnetization plateau at $M=1/3\,$. An
additional feature, particular to the Kagom\'e lattice is that the
UUD state is not a N\'eel ordered state, rather it is a spin
liquid. Our evidence for this is that we find a number of states
with comparable weights.
It seems therefore that the strong fluctuations associated with
the the spin liquid state do not destroy the plateau; rather, the
two effects co-habit. Further investigation of these issues is
underway. At the classical level a plateau  occurs in a region of
($\Delta, \; h$) phase space, terminating at the point ($\Delta =
1,\; h=2JS$), where a pseudo plateau is driven by the entropic
contribution to the free energy. The results of this paper allow
us to speculate that quantum fluctuations stabilize a plateau over a
more extensive region of parameter space. The robustness of this
result appears to be associated with a clean separation of the GS
components according to their weight scale, the highest of which
are UUD configurations.

\vskip 0.15cm

We aknowledge useful discussions with J.T. Chalker, D. Champion,
P. Degiovanni, F. Delduc, E.\ Fradkin, L.\ Freidel, A.\ Honecker,
O. Petrenko, G.L.\ Rossini and M.\ Zhitomirsky and the financial
support of the ECOS-Sud committee. The research of D.C.C and
M.D.G. is partially supported by CONICET and Fundaci\'on
Antorchas, Argentina (grant No.\ A-13622/1-106).



\begin{thebibliography}{99}
\bibitem{boundaries} Note that periodic boundaries are
essential; free boundary effects yield neither evidence of massive
spin excitations nor of sharp weight distributions.
\end{thebibliography}
\end{document}